\documentclass{aastex}
\usepackage[utf8]{inputenc}
\usepackage{natbib}
\usepackage{amsthm}
\usepackage{amstext}
\usepackage{amsmath}
\usepackage{amsbsy}
\usepackage{amssymb}
\usepackage{amsfonts}
\usepackage{graphicx}
\usepackage{epsfig}
\usepackage{hyperref}
\usepackage{subfigure}
\usepackage{textcomp}

\shorttitle{Anisotropies in cosmic-ray positrons measured by PAMELA}
\shortauthors{Adriani et al.}

\begin{document}
\title{ Search for anisotropies in cosmic-ray positrons detected by the PAMELA experiment }

\author{
O.~Adriani$^{1,2}$, G.~C.~Barbarino$^{3,4}$, G.~A.~Bazilevskaya$^{5}$, R.~Bellotti$^{6,7}$, M.~Boezio$^{8}$, E.~A.~Bogomolov$^{9}$, M.~Bongi$^{1,2}$,
V.~Bonvicini$^{8}$, S.~Bottai$^{2}$, A.~Bruno$^{6}$, F.~Cafagna$^{7}$, D.~Campana$^{4}$, P.~Carlson$^{10}$,
M.~Casolino$^{11,12}$,
G.~Castellini$^{13}$, C.~De Donato$^{11,14}$, C.~De Santis$^{11,14}$, N.~De Simone$^{11}$,
V.~Di Felice$^{11,15}$, V.~Formato$^{8,16}$, A.~M.~Galper$^{17}$,
U.~Giaccari$^{4,20}$, A.~V.~Karelin$^{17}$, S.~V.~Koldashov$^{17}$, S.~Koldobskiy$^{17}$,
S.~Y.~Krutkov$^{9}$, A.~N.~Kvashnin$^{5}$, A.~Leonov$^{17}$,
V.~Malakhov$^{17}$, L.~Marcelli$^{11,14}$,
M.~Martucci$^{14,18}$, A.~G.~Mayorov$^{17}$, W.~Menn$^{19}$, M.~Merg$\acute{e}$ $^{11,14}$, V.~V.~Mikhailov$^{17}$,
E.~Mocchiutti$^{8}$, A.~Monaco$^{6,7}$, N.~Mori$^{1,2}$, R.~Munini$^{8,16}$, G.~Osteria$^{4}$, F.~Palma$^{11,14}$, B.~Panico$^{4,*}$, P.~Papini$^{2}$,
M.~Pearce$^{10}$, P.~Picozza$^{11,14}$, M.~Ricci$^{18}$, S.~B.~Ricciarini$^{2,13}$, R.~Sarkar$^{8,21}$, V.~Scotti$^{3,4}$, M.~Simon$^{19}$,
R.~Sparvoli$^{11,14}$, P.~Spillantini$^{1,2}$, Y.~I.~Stozhkov$^{5}$, A.~Vacchi$^{8}$, E.~Vannuccini$^{2}$, G.~I.~Vasilyev$^{9}$, S.~A.~Voronov$^{17}$,
Y.~T.~Yurkin$^{17}$, G.~Zampa$^{8}$, and N.~Zampa$^{8}$}

\vspace{5cm}
\affil{$^1$University of Florence, Department of Physics and Astronomy, I-50019 Sesto Fiorentino,Florence, Italy}
\affil{$^2$INFN, Sezione di Firenze, I-50019 Sesto Fiorentino, Florence, Italy}
\affil{$^3$University of Naples `Federico II', Department of Physics, I-80126 Naples, Italy}
\affil{$^4$INFN, Sezione di Napoli, I-80126 Naples, Italy}
\affil{$^5$Lebedev Physical Institute, RU-119991, Moscow, Russia}
\affil{$^6$University of Bari, Department of Physics, I-70126 Bari, Italy}
\affil{$^7$INFN, Sezione di Bari, I-70126 Bari, Italy}
\affil{$^8$INFN, Sezione di Trieste, I-34149 Trieste, Italy}
\affil{$^9$Ioffe Physical Technical Institute, RU-194021 St. Petersburg, Russia}
\affil{$^{10}$KTH, Department of Physics, and the Oskar Klein Centre for Cosmoparticle Physics, AlbaNova University Centre, SE-10691 Stockholm, Sweden}
\affil{$^{11}$INFN, Sezione di Roma Tor Vergata, I-00133 Rome, Italy}
\affil{$^{12}$RIKEN, Advanced Science Institute, Wako-shi, Saitama, Japan}
\affil{$^{13}$IFAC, I-50019 Sesto Fiorentino, Florence, Italy}
\affil{$^{14}$University of Rome Tor Vergata, Department of Physics, I-00133 Rome, Italy}
\affil{$^{15}$Agenzia Spaziale Italiana (ASI) Science Data Center, Via del Politecnico snc I-00133 Rome, Italy}
\affil{$^{16}$University of Trieste, Department of Physics, I-34147 Trieste, Italy}
\affil{$^{17}$National Research Nuclear University MEPhI, RU-115409 Moscow}
\affil{$^{18}$INFN, Laboratori Nazionali di Frascati, Via Enrico Fermi 40, I-00044 Frascati, Italy}
\affil{$^{19}$Universitat Siegen, Department of Physics, D-57068 Siegen, Germany}
\affil{$^{20}$Now at Universidade Federal do Rio de Janeiro, Instituto de Fisica, Rio de Janeiro, RJ, Brazil}
\affil{$^{21}$Previously at INFN, Sezione di Trieste, I-34149 Trieste, Italy}

\altaffiltext{*}{Corresponding author. E-mail address: beatrice.panico@na.infn.it.}

\begin{abstract}
The PAMELA detector was launched on board of the Russian Resurs-DK1 satellite on June 15, 2006. Data collected during the first four years 
have been used to search for
large-scale anisotropies in the arrival directions of
cosmic-ray positrons. The PAMELA experiment allows for a full sky investigation,
with sensitivity to global anisotropies in any angular window of the celestial sphere.
Data samples of positrons in the rigidity range 10 GV $\leq$ R $\leq$ 200 GV were analyzed.
This article discusses the method and the results of the search for possible local sources through analysis of anisotropy
in positron data compared to the proton background.
The resulting distributions of arrival directions are found to be isotropic. 
Starting from the angular power spectrum, a dipole anisotropy upper limit $\delta$ = 0.166 at 95\% C.L. is determined.
Additional search is carried out around the Sun. No evidence of an excess correlated with that direction was found.
\end{abstract}

\keywords{PAMELA, cosmic ray, anisotropy, positron}

\section{Introduction}

Measurements of cosmic-ray positrons address a number of questions in contemporary astrophysics, such as the nature and distribution of particle sources in our Galaxy, and the subsequent
propagation of cosmic-rays through the Galaxy and the solar magnetosphere. 
Positrons are a natural component of the cosmic radiation, produced in the interaction between cosmic rays and the interstellar matter. 
Observations by the PAMELA experiment~\citep{Adriani09,Adriani10,Adriani13}, confirmed by other recent measurements \citep{Ackermann12,Accardo14},
revealed a positron excess over the predictions of commonly used propagation models, e.g. \citep{Moskalenko09, Delahaye09}.
Either the standard paradigm of cosmic-ray propagation in the galaxy should be revised or additional sources of cosmic-ray should be introduced to explain the
positron anomaly at high energy. 
There are two primary hypotheses for these new sources:
1) a source linked to particle physics, e.g. a dark matter decay or
annihilation, e.g. \citep{Cirelli08}.
2) a nearby astrophysical source, e.g. a pulsar or supernova remnant,
e.g. \citep{Hooper09}.

In the first case, among the various dark matter models proposed,
\citet{Batell10}, \citet{Schuster10} suggested that a significant fraction of
positrons detected at Earth could be produced by dark matter particles
annihilating in the neighborhood of the Sun. 
In this case, the electron/positron pair emission would be strongly
correlated with the center 
of the Sun and would produce visible effect in anisotropy study.

In the second case, the localization of the astrophysical sources 
could results in anisotropy in the arrival direction of the cosmic-ray
positrons.
Several authors, e.g. \citep{Ptuskin95, Kobayashi04, Bushing08},
estimated that the expected anisotropy in cosmic-ray positrons from
supernova 
remnants and/or from pulsars should be of the order of percent or less. 

Directional and timing data from PAMELA are exploited to search for positron anisotropies that can provide unique information on new sources of cosmic rays.
Previous searches have been already carried out using the combined flux of electrons and positrons \citep{Ackermann12,Accardo14, Campana13} and results were
compatible with an isotropic distribution of arrival directions of detected particles.

In this paper the analysis is focused on positrons since they represent a cleaner sample of cosmic rays produced by possible sources respect to electrons.
In most theoretical models, electrons and positrons from new sources are produced in pairs. However the e$^+$ flux consists of only two components: e$^+$ from 
secondary production and from unknown sources, while the e$^-$ flux contains, in addition and in a larger amount, also primary e$^-$.
Since in this energy range the cosmic-ray primary electrons represent the 90\% of the total e$^-$ flux,
the search for anisotropies in the pure electron sample will be dominated by an isotropic background. 

In the following, an analysis method is presented to study the anisotropy of positrons using the back-tracing of particle trajectories,
to identify arrival directions far from the Earth. An evaluation of systematic uncertainties is also provided.

%%%%%%%%%%%%%%%%%%%%%%%%%%%%%%%%%%%%%%%%%%%%%%%%%%%%%%%%%%%%%%%%%%%%%%%%%%%%%%%%%%%%%%%%%%%%%%%%%%%%%%%%%%%%%%%%%%%%%%%%%%%%%%%%%%%%%%%%%%%%%%%%%%%%%%
\section{The PAMELA detector}

The PAMELA experiment has been operating on board of the satellite Resurs-DK1 since June 2006. 
The design is optimized for precision studies of 
light particles and antiparticles in primary cosmic rays between a few tens of MeV and a few hundred of GeV.

Since launch, PAMELA has collected $\sim 8 \cdot 10^9$ events, of which $\sim 5 \cdot 10^6$ electrons and positrons, whose analysis is described
in several publications \citep{Adriani09,Adriani10,Adriani11}.

The PAMELA detector \citep{Picozza07} is equipped with a magnetic spectrometer, comprising a permanent magnet
hosting a tracking system. The tracking system consists of six double-sided microstrip silicon planes, which
allow the determination of the particle charge and rigidity (momentum divided by charge) with high precision.
An imaging electromagnetic calorimeter, made of 44 silicon planes interleaved with 22 plates of tungsten
absorber, is mounted below the spectrometer to accurately perform particle identification and to measure the energy of electrons and protons.
A Time of Flight (ToF) system made of three planes of double layers of plastic scintillator permits measurements of
particle velocity and energy loss. It also provides the main trigger for the experiment.
Another layer of scintillator and a neutron detector, placed below the calorimeter, give additional information about the shower extent and
improves lepton/hadron discrimination.
Particles in the PAMELA acceptance due to scattering or interactions are rejected during off-line analysis of the anti-coincidence system signals.

This apparatus permits electrons and positrons to be separated from the proton background over the rigidity range 10 GV $\leq$ R $\leq$ 200 GV and to
measure their incoming direction with an accuracy of about two degrees. For each particle crossing
the detector the arrival direction from space is reconstructed using the trajectory inside the
instrument and the satellite position. The satellite orbit, 70$^{\circ}$ inclination and
350-610 km altitude, allows PAMELA to perform a very detailed measurement of the cosmic radiation in different
regions of Earth's magnetosphere. Back tracing of detected particles in the geomagnetic field is
performed in order to obtain their initial spatial distribution, outside of the Earth magnetosphere.

%%%%%%%%%%%%%%%%%%%%%%%%%%%%%%%%%%%%%%%%%%%%%%%%%%%%%%%%%%%%%%%%%%%%%%%%%%%%%%%%%%%%%%%%%%%%%%%%%%%%%%%%%%%%%%%%%%%%%%%%%%%%%%%%%%%%%%%%%%%%%%%%%%%%%%
\section{Analysis Method}
\label{Analysis}

This data analysis aims to identify the presence of any large-scale pattern in the distribution of arrival directions
of cosmic-ray positrons detected by PAMELA.
The entire data-set of PAMELA up to January 2010 was analyzed, constructing sky maps for positrons in Galactic coordinates.
In order to perform a search for anisotropies in these data, the experimental particle flux,
measured by the instrument for different directions, is compared with background (or coverage) maps derived from an isotropic particle flux.

Coverage maps can be determined through an accurate simulation of particles coming from
any direction of the sky, but this would require a very precise knowledge of
the instrument exposure, defined as the projection of the acceptance in
each sky direction integrated over the full live time of the detector.
Since the result depends on the position and orientation of the spacecraft as a function of time, small inaccuracies 
can lead to the creation of fake anisotropy signals.
To overcome this problem, the background map for isotropic expectation was constructed using
PAMELA data themselves. The flux of protons is highly isotropic, as previously shown with PAMELA data \citep{Giaccari13}. 
A set of proton events was selected as described in Sec.~\ref{selection} for the same 
period of time as used for the positron data.
The instrument exposure, dead times and other detector effects are therefore included when calculating the ratio of
positrons over protons.
The comparison between signal and background maps was performed with two independent and complementary techniques: 
a statistical significance test introduced by Li and Ma \citep{Li83} and a
spherical harmonic analysis.

\subsection{Back tracing}

In order to identify particle arrival directions and investigate possible anisotropies, it is necessary to account for the effect of the
geomagnetic field on particle propagation.
Consequently, trajectories of all selected events were reconstructed in
the Earth's magnetosphere by means of a tracing program \citep{Bruno14}
based on numerical integration methods \citep{Smart00}.
The IGRF-11 \citep{Finlay10} model and the TS05 \citep{Tsyganenko05} model were used for the description of the geomagnetic field:
the former employs a global spherical harmonic implementation of the
main magnetic field; the latter provides a dynamical (five-minute resolution), semi-empirical
best-fit representation of the external geomagnetic field sources in the
inner magnetosphere, based on recent satellite measurements.
Using satellite ephemeris (position, orientation, time information) and the
particle rigidity and direction measured by the tracking system,
trajectories were back-propagated to the model magnetosphere boundaries,
allowing corresponding asymptotic arrival directions (i.e. the
directions of approach before they encountered the geomagnetic field) to
be determined.

\subsection{Sky Pixelation}

To determine the arrival direction distribution over the whole sky, the Healpix framework was used. Healpix provides a visualization of the sky map in a 2D-sphere \citep{Gorski05}.
This tool pixelises (i.e. a subdivides) the 2D-sphere, where each pixel covers the same surface area as every other pixel and is regularly distributed
on lines of constant latitudes. In the Healpix scheme the pixel size cannot be chosen arbitrarily, the total number of pixels is given by $12 \times n_{side}^{2}$ where $n_{side}$
defines the map resolution and can take only a value that is power of two.
A grid with 3072 pixels with an angular pixel extension of about $\sim~7^{\circ}$ was adopted. All maps are given in Galactic coordinates.

\subsection{Significance Maps}
\label{SigMap}

The statistical significance of the signal over an isotropic background in
each direction of the sky was derived following the technique described in the work of Li and Ma \citep{Li83}.
The significance formula is defined as
\begin{equation}
 \begin{split}
S= \pm \sqrt{2} \Bigg\{ & N_{\textbf{on}}\ln\left[\frac{1+\alpha}{\alpha}\left(\frac{N_{\textbf{on}}}{N_{\textbf{on}}+N_{\textbf{off}}}\right)\right]+ \\
& N_{\textbf{off}}\ln\left[(1+\alpha)\left(\frac{N_{\textbf{off}}}{N_{\textbf{on}}+N_{\textbf{off}}}\right)\right] \Bigg\}^{1/2},
\end{split}
\label{Li-MaEq}
\end{equation}
where $N_{\textbf{on}}$ and $N_{\textbf{off}}$ are respectively the observed and the expected
number of events in a certain angular window of the sky,
after the proton and positron maps are normalized to the same number of events.

In Eq.\eqref{Li-MaEq} $\alpha$ is
the ratio of the exposure in the \emph{on-source} region divided by the exposure in the \emph{off-source}
region; in this case the \emph{on-source} and \emph{off-source} regions coincide so $\alpha = 1$. \\
The significance is defined as positive if there is an excess
otherwise as negative in the presence of a deficit of events.
If no signal is measured, the distribution of the significance $S$ is a
normal function. Results are reported bin per bin, as a number of standard
deviations.

This technique, developed in $\gamma$-astronomy to search for point sources, can
be easily arranged to search large anisotropies integrating the signal and the background maps over
various angular scales in order to increase the sensitivity up to a large-scale pattern.
In this survey different angular radii for the integration, from 10$^{\circ}$ to 90$^{\circ}$, were used.
The center of each bin of the sky pixelization is considered and the content of
the bin is given by the cumulative number of events falling within a given angular distance from the
center of that bin. Both signal and background maps are smoothed following this process. In this way
the obtained integrated maps have correlated neighboring pixels and none of the information at a given
angular scale is lost.

The statistical Li and Ma significance test is subsequently applied to study possible deviations from isotropy.
In the case of no strong anisotropies, each significance histogram approximates a Gaussian
distribution.
The significance distributions becomes narrower with increasing integration radius - this is expected as in the integrated maps the bins are
strongly correlated and identical events are included in the calculation.

\subsection{Spherical Harmonic Analysis}

The orbital parameters of the Resurs-DK1 satellite provide a relatively uniform exposure over the full
celestial sphere. This permits the study of the angular power spectrum of arrival directions of the events, 
i.e. the cosmic-ray intensity over the celestial sphere can be expanded in spherical harmonics.
The power spectrum analysis is a powerful tool to study the anisotropy patterns and provides information on the angular scale of the anisotropy in the map.

The positron (or electron) relative intensity map $I$, defined as the relative deviation of the
observed number of events $N$ from the expected number of events $<N>$ in each angular bin of the sky defined 
by the Galactic coordinates $(glon,glat)$ is considered, 
\begin{equation}
\label{eq:Fluct}
I(glon,glat)=\frac{N(glon,glat)-<N(glon,glat)>}{<N(glon,glat)>}.
\end{equation}
\noindent The relative map $I$ is expanded in the basis of spherical harmonics functions $Y_{lm}$
\begin{equation}
I(glon,glat)=\sum_{l=0}^{\infty}\sum_{m=-l}^{m=l} a_{lm}Y_{lm}(glon,glat).
\end{equation}
The full anisotropy information is encoded into the set of spherical harmonic coefficients, $a_{lm}$, which can be used to calculate the angular power spectrum
\begin{equation}
 C(l)=\frac{1}{2l+1}\sum_{m = -l}^{m = l} a^{2}_{lm}.
 \label{eq:Cl}
\end{equation}
The amplitudes of the power spectrum at some multipole order are associated to the presence of structures in the distribution of the arrival directions in the sky.
Non-zero amplitudes in the multipole moments, $C_{l}$, arise from fluctuations in the particle flux on an angular scale near~$180/l$~degrees.

\subsubsection{Dipole upper limit}
\label{dip_ul}
Although all the spherical harmonic moments are important to characterize the anisotropy patterns at any angular scale, 
the dipole moment is of particular interest \citep{Bushing08}.
In fact, if there is a marked directionality, dipole anisotropy is expected as could be the case for a single source dominating the positron flux.
In this case the overall intensity at an angular distance $\theta$ from the maximum of the dipole anisotropy can 
be written as I($\theta$)=I$_0$+I$_1$cos($\theta$), where I$_0$ represents the isotropic signal and 
I$_1$ represents the maximum anisotropy due to the dipole. 
The degree of anisotropy can be expressed as the fraction $\delta=I_1/I_0$.

By applying Eq. \ref{eq:Fluct} the intensity of the dipole anisotropy becomes
\begin{equation}
I(\theta)=\frac{I(\theta)-<I(\theta)>}{<I(\theta)>}=\frac{I(\theta)-I_0}{I_0}=\frac{I_1cos(\theta)}{I_0}.
\end{equation}
Since the spherical harmonic $Y_1^0(\theta,\phi)$ depends from $\theta$, it is possible to derive that $a_1^0=\frac{I_1}{I_0}\sqrt{\frac{4\pi}{3}}$ and, 
consequently, 
\begin{equation}
 C_1=\Bigl(\frac{I_1}{I_0}\Bigl)^2 \frac{4\pi}{9}.
\label{eq:Cl2}
 \end{equation}
Starting from Eq. \ref{eq:Cl2}, the correlation between the dipole amplitude and the power spectrum coefficient with $l = 1$ is
\begin{equation}
\delta = 3 \sqrt{\frac{C_{1}}{4 \pi}}.
\label{eq:aps}
\end{equation}
Considering $\hat{C_{1}}$ as the outcome of the calculated angular power spectrum over the signal map for \emph{l}=1, for the observed spectrum 
\begin{equation}
\hat{\delta} = 3 \sqrt{\frac{\hat{C_{1}}}{4 \pi}}.
\label{eq:DeltaHat}
\end{equation}
Since $\hat{C_{1}}$ follows a $\chi^2$ distribution with three degrees of freedom centered on $C_1$, the probability to measure $\hat{C_{1}}$ given the true $C_1$
is 
\begin{equation}
P(\hat{C_1};C_1)\frac{3\sqrt{3}}{\sqrt{2\pi}C_1}\sqrt{\frac{\hat{C_1}}{C_1}}~exp\Bigl(- \frac{3\hat{C_1}}{2C_1}\Bigl).
\label{eq:ProbC1}
\end{equation}
From Eq. \ref{eq:DeltaHat} and \ref{eq:ProbC1} it is possible to derive the probability for $\hat{\delta}$ as 
\begin{equation}
P(\hat{\delta};\delta)=\frac{3 \sqrt{6}}{\sqrt{\pi{}}} \frac{\hat{\delta}^2}{\delta^3} ~ exp\Bigl(-\frac{3\hat{\delta}^2}{2\delta^2}\Bigl).
\end{equation}
The upper limit on $\delta$ can be calculated using a frequentist approach where the limit corresponds to the solution of the integral 
\begin{equation}
%\int_{0}^{\hat{\delta}} P(\hat{\delta} \vert \delta) ~ d\delta = 1 - C.L.
\int_{0}^{\hat{\delta}_{\textrm{m}}} P(\hat{\delta} \vert \delta) ~ d\hat{\delta} = 1 - C.L.
\label{eq:ul}
\end{equation}
for a given confidence level (C.L.) \citep{Neyman37,Garthwaite95}, where $\hat{\delta}_{\textrm{m}}$ represents the dipole anisotropy corresponding to the 
measurement of the dipole power spectrum $\hat{C}_{1,m}$.

%%%%%%%%%%%%%%%%%%%%%%%%%%%%%%%%%%%%%%%%%%%%%%%%%%%%%%%%%%%%%%%%%%%%%%%%%%%%%%%%%%%%%%%%%%%%%%%%%%%%%%%%%%%%%%%%%%%%%%%%%%%%%%%%%%%%%%%%%%%%%%%%%%%%%%
\section{Data Selection}
\label{selection}

The sample of positrons selected for this analysis concerns to the period from the satellite launch
to the end of the solar minimum phase (June 2006 - January 2010). 
This is the same data set as used for previous publications \citep{Adriani10}. The selection criteria are, however, different 
since a statistical subtraction of proton contamination in the
positron sample does not allow event-by-event positron identification as required to determine the 
incoming direction and rigidity of each event.
The positron sample was instead selected using a cut based analysis similar to the one developed for the electron
flux determination \citep{Adriani11}. The same track and event quality selection were used; the hadron
contamination was reduced to a negligible amount by using a stronger calorimeter selection
(with about 80\% selection efficiency) combined with activity requirements on the neutron counter.
A very clean sample of positrons was obtained from flight data in the rigidity range from 10 to 200 GV.
A total number of 1489 positrons with rigidity R $>$ 10 GV was selected for this work.
With the same selection a sample of 20673 electrons and positrons was selected for the anisotropy search in the direction of the Sun.
The quality of the sample was verified by comparing the resulting positron fraction to the published one.
In the same period, June 2006 to January 2010,
the calorimeter was used to select hadronic showers initiated by positively charged particles
in the same rigidity range and with the same requirements on the event and track quality.
In this way a proton sample of about $4.5 \cdot 10^5$ events,
which preserved the angular distribution of particles along PAMELA orbit, was obtained.
Finally, a refined analysis on the detector orientation was carried out by comparing the pointing of the instrument determined
by the satellite orbital information with the pointing obtained using data from the Earth remote sensing camera that is installed on board the Resurs-DK1 satellite.
This refined analysis minimized the uncertainty on the measured absolute incoming trajectory for all the
selected particles which is used in the back-tracing and in the anisotropy maps. \\

%%%%%%%%%%%%%%%%%%%%%%%%%%%%%%%%%%%%%%%%%%%%%%%%%%%%%%%%%%%%%%%%%%%%%%%%%%%%%%%%%%%%%%%%%%%%%%%%%%%%%%%%%%%%%%%%%%%%%%%%%%%%%%%%%%%%%%%%%%%%%%%%%%%%%%

\section{Results}

%%%%%%%%%%%%%%%%%%%%%%%%%%%%%%%%%%%%%%%%%%%%%%%%%%%%%%%%%
\begin{figure}[h!]
\centering
\subfigure[\protect\url{Positrons}]%
{\includegraphics[width=8cm]{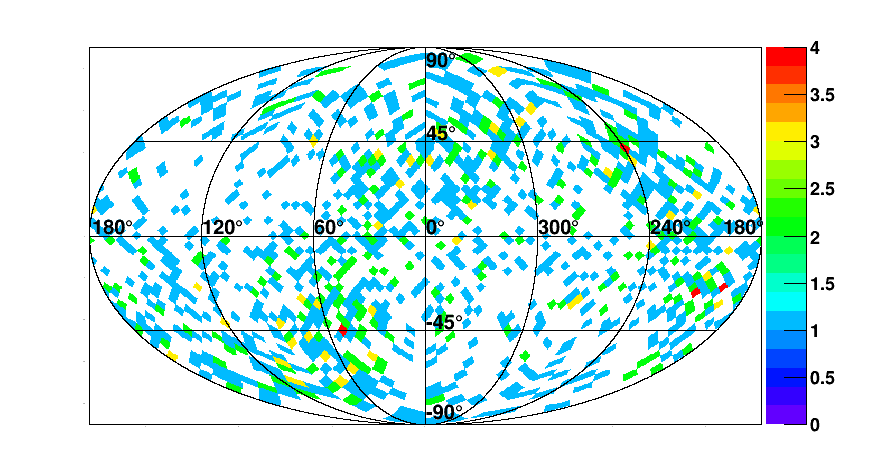}}
\subfigure[\protect\url{Protons}]%
{\includegraphics[width=8cm]{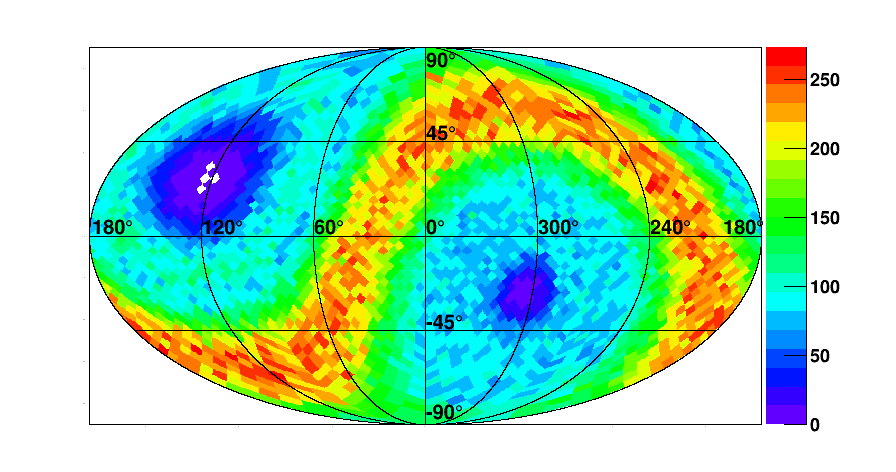}}
\caption[Event and the coverage maps]{Event maps for positrons (a) and protons (b)
for 10 GV $\leq$ R $\leq$ 200 GV, taking into account the geomagnetic field effect.}
\label{fig:map}
\end{figure}
\begin{figure}[h!]
\centering
\includegraphics[width=8cm]{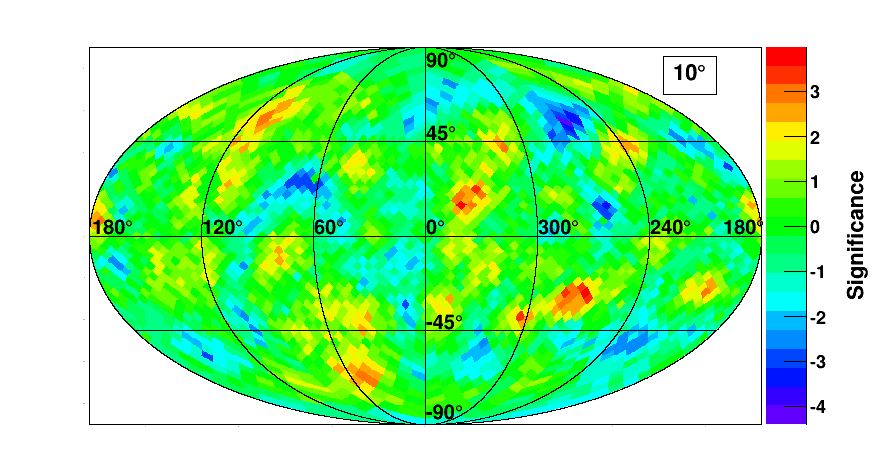}
\includegraphics[width=8cm]{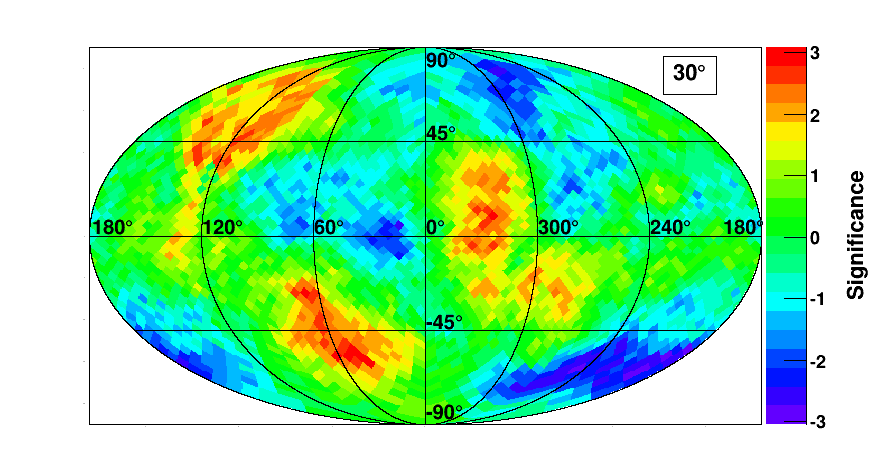}
\includegraphics[width=8cm]{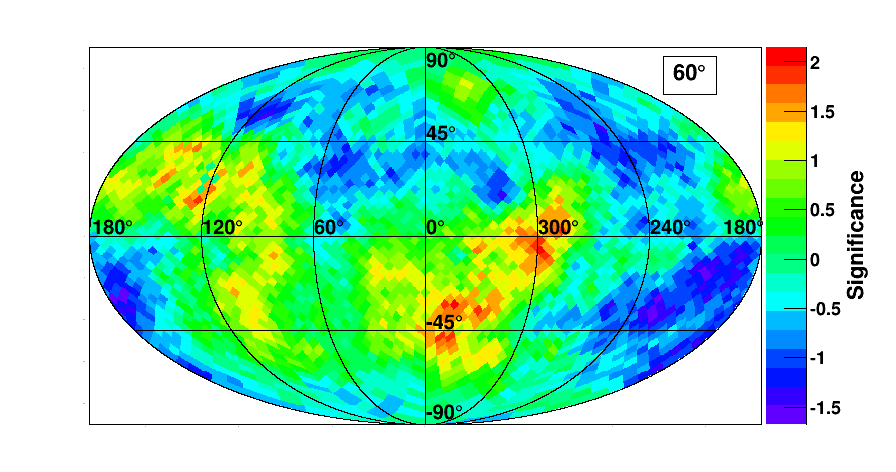}
\includegraphics[width=8cm]{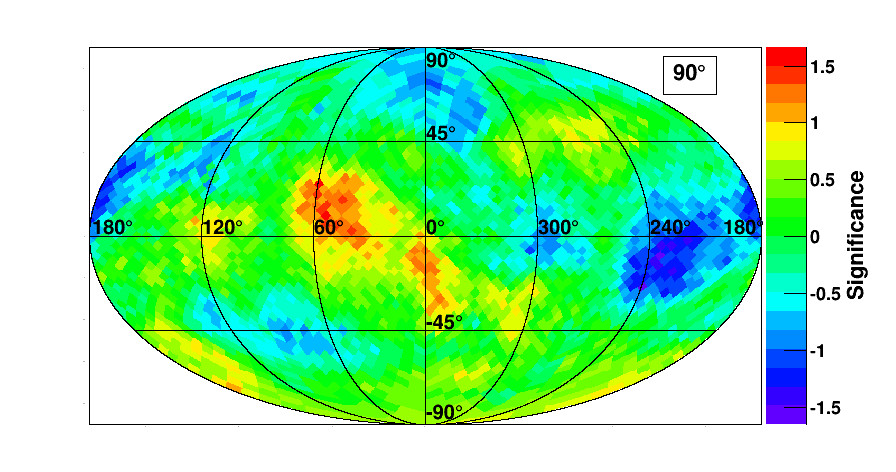}
\caption[LiMa maps]{Significance maps
for 10 GV $\leq$ R $\leq$ 200 GV, over the following angular scales: 10$^{\circ}$, 30$^{\circ}$, 60$^{\circ}$, 90$^{\circ}$.}
\label{fig:LiMa}
\end{figure}

%%%%%%%%%%%%%%%%%%%%%%%%%%%%%%%%%%%%%%%%%%%%%%%%%%%%%%%%%
Fig.~\ref{fig:map} shows the maps obtained for events with rigidity greater than 10 GV,
for selected positrons and protons, taking into account the geomagnetic field.
As the PAMELA experiment has a shorter exposure to the poles, the sky is not observed uniformly, leading to a particular
shape in the observed maps. The same overall exposure is achieved for both types of particle.
Both the signal and the background maps were binned taking into account the detector angular resolution. The color scale represents 
the number of detected events for each bin. If a signal can be found in the data, it could spread over multiple adjacent bins, 
making the anisotropy measurement more difficult. Hence, to highlight a possible anisotropy, the bin size must be similar to 
the angular scale of the anisotropy itself. To increase sensitivity it is possible to merge the content of neighbouring bins. 
The content of each integrated bin is equal to the total number of events belonging to the bins covered by a circular region 
of a given radius. The radius chosen for the integration represents the angular scale at which the anisotropy is expected. 
If the integration radius is too small the signal is distributed among adjacent bins, while if it is too large too 
much background will be integrated.
The events and the coverage maps were integrated over the following angular scales: 10$^{\circ}$, 30$^{\circ}$, 60$^{\circ}$, 90$^{\circ}$.
%%%%%%%%%%%%%%%%%%%%%%%%%%%%%%%%%%%%%%%%%%%%%%%%%%%%%%%%%
\begin{figure}[htb!]
\centering
\includegraphics[width=8cm]{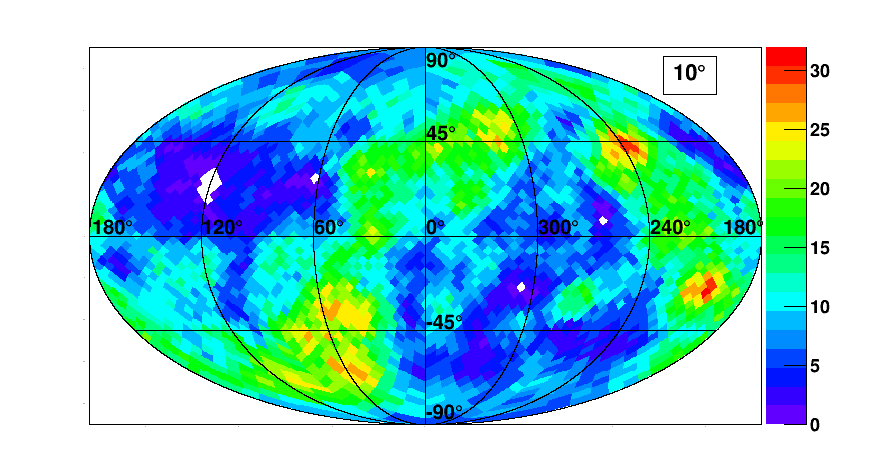}
\includegraphics[width=8cm]{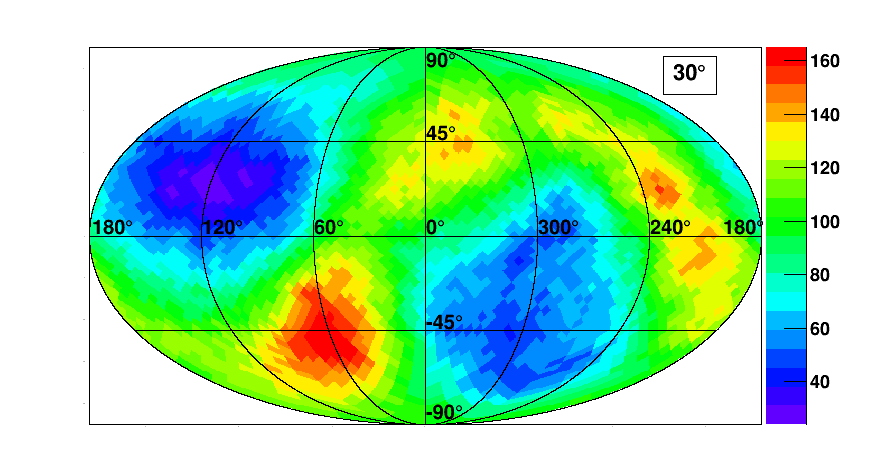}
\includegraphics[width=8cm]{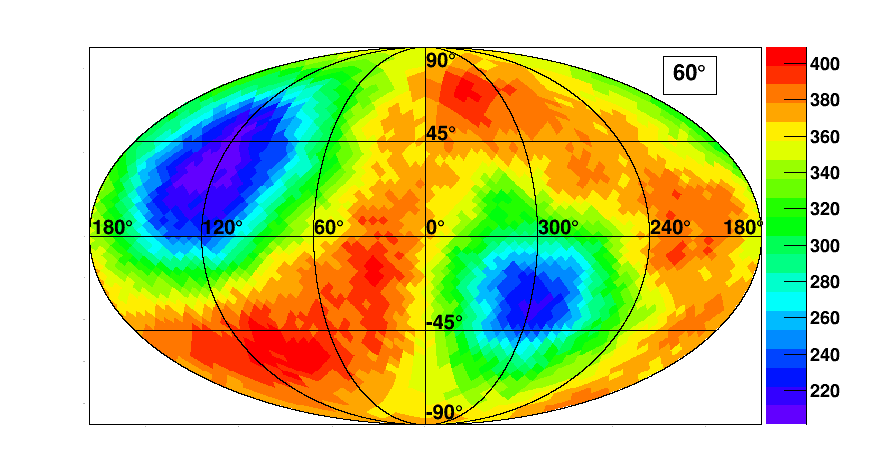}
\includegraphics[width=8cm]{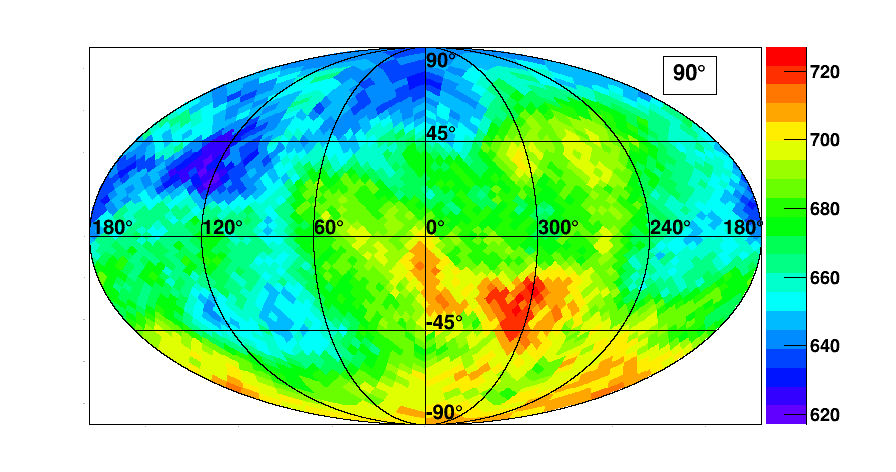}
\caption[Different integration radii for positron map]{Positron maps
for 10 GV $\leq$ R $\leq$ 200 GV, over the following angular scales: 10$^{\circ}$, 30$^{\circ}$, 60$^{\circ}$, 90$^{\circ}$.}
\label{fig:intpos}
\end{figure}
%%%%%%%%%%%%%%%%%%%%%%%%%%%%%%%%%%%%%%%%%%%%%%%%%%%%%%%%%
%%%%%%%%%%%%%%%%%%%%%%%%%%%%%%%%%%%%%%%%%%%%%%%%%%%%%%%%%
\begin{figure}[htb!]
\centering
\includegraphics[width=8cm]{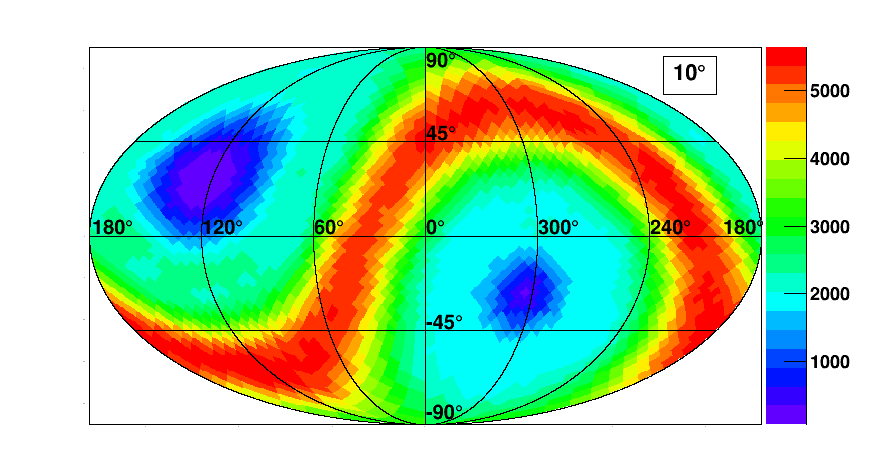}
\includegraphics[width=8cm]{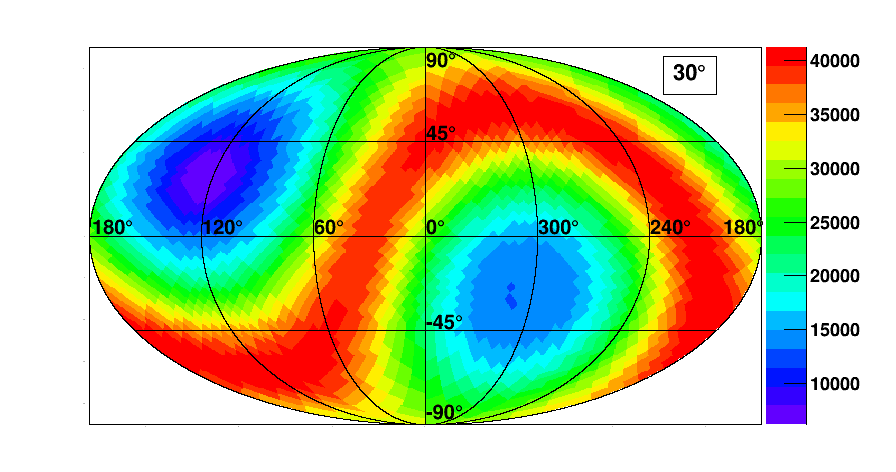}
\includegraphics[width=8cm]{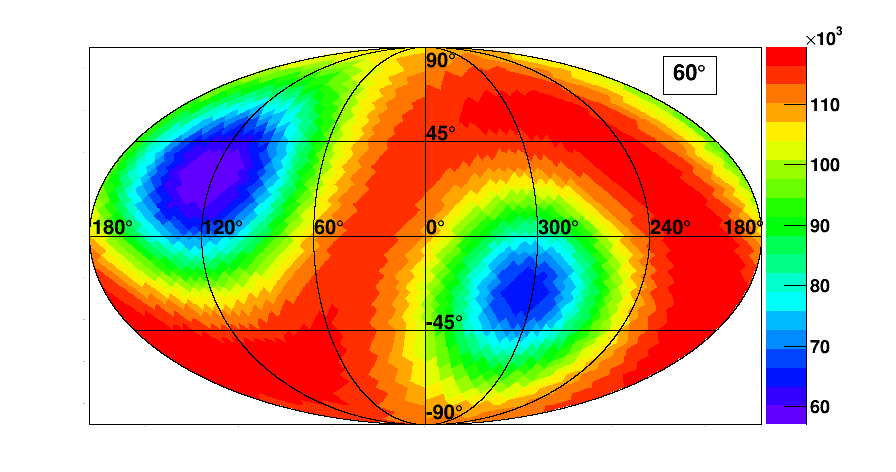}
\includegraphics[width=8cm]{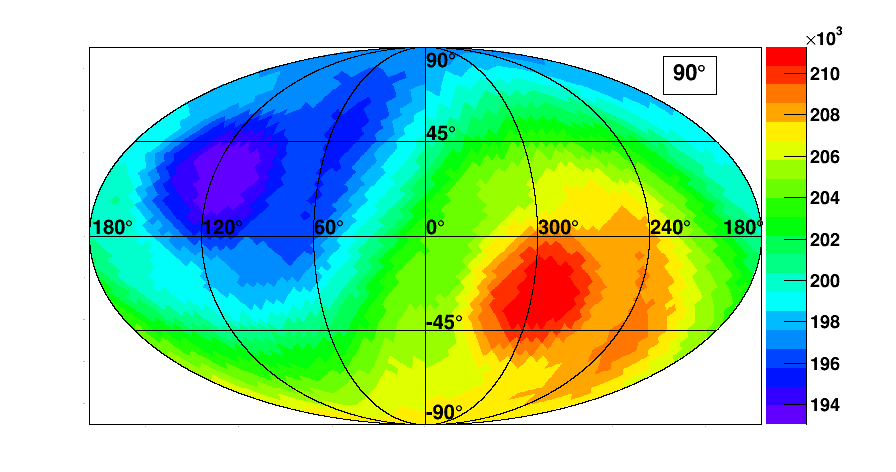}
\caption[Different integration radii for proton map]{Proton maps
for 10 GV $\leq$ R $\leq$ 200 GV, over the following angular scales: 10$^{\circ}$, 30$^{\circ}$, 60$^{\circ}$, 90$^{\circ}$.}
\label{fig:intprot}
\end{figure}
%%%%%%%%%%%%%%%%%%%%%%%%%%%%%%%%%%%%%%%%%%%%%%%%%%%%%%%%%
The corresponding results are reported in Fig.~\ref{fig:intpos} for positrons and in Fig.~\ref{fig:intprot} for protons. 
The color scale now represents the integrated number of events. The exposure distribution varies depending on the integration radius as expected.
To estimate the statistical significance of any excess or deficit the method detailed 
in Sec.~\ref{SigMap} was used. In this method, likelihood functions are applied both to the null hypothesis and to the signal one;
the ratio between these two functions represents the significance of the excess.
Significance maps, constructed by comparing signal and corresponding background maps for the different integration radii, are
shown in Fig.~\ref{fig:LiMa}. The significance signals are always less than four sigma, indicating that a point-like source 
cannot be distinguished from background.
In the case of statistical fluctuations for an isotropic sky the distribution of the significance will exhibit a Gaussian distribution with zero mean.
%%%%%%%%%%%%%%%%%%%%%%%%%%%%%%%%%%%%%%%%%%%%%%%%%%%%%%%%%
%%%%%%%%%%%%%%%%%%%%%%%%%%%%%%%%%%%%%%%%%%%%%%%%%%%%%%%%%
%%%%%%%%%%%%%%%%%%%%%%%%%%%%%%%%%%%%%%%%%%%%%%%%%%%%%%%%%
\begin{figure}[h!]
\centering
\subfigure[\protect\url{}]
{\includegraphics[width=8cm]{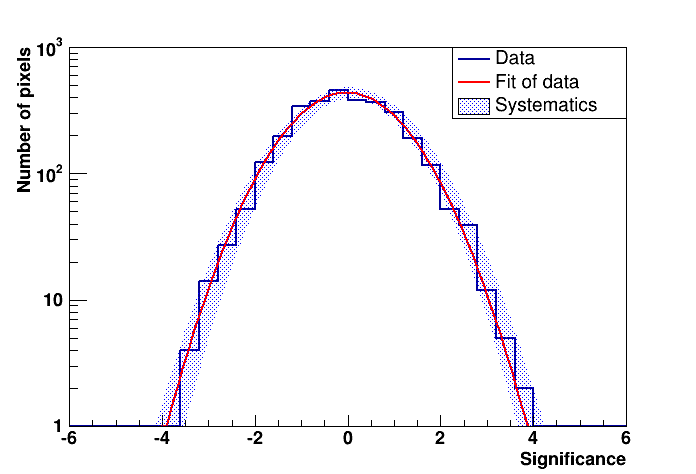}}
\subfigure[\protect\url{}]
{\includegraphics[width=8cm]{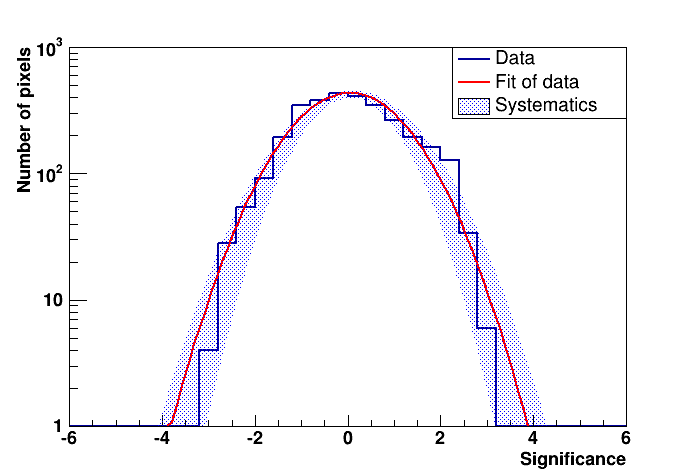}}
\subfigure[\protect\url{}]
{\includegraphics[width=8cm]{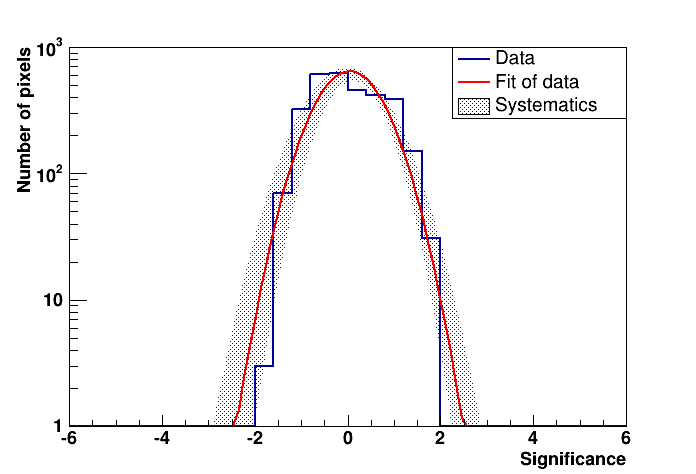}}
\subfigure[\protect\url{}]
{\includegraphics[width=8cm]{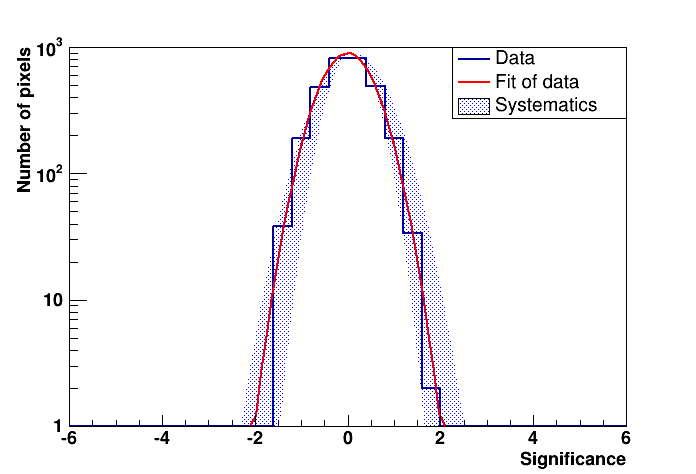}}
\caption[LiMa histograms]{Histograms for the significance maps
for 10 GV $\leq$ R $\leq$ 200 GV, over the following angular scales: 10$^{\circ}$ (a), 30$^{\circ}$ (b), 60$^{\circ}$ (c), 90$^{\circ}$ (d).
The grey area represents the systematics.}
\label{fig:histoLiMa}
\end{figure}
%%%%%%%%%%%%%%%%%%%%%%%%%%%%%%%%%%%%%%%%%%%%%%%%%%%%%%%%%
In Fig.~\ref{fig:histoLiMa}, the obtained significance distributions are shown together with a fitted Gaussian function; no significant
deviation from isotropy is observed.
It can be seen that the significance distribution becomes narrower when increasing the integration radius since with
larger integration radius, the bins are strongly correlated, reducing the variance of the distribution.
For the study of the angular power spectrum of the arrival direction distribution, the cosmic-ray intensity was expanded
in spherical harmonics, using the \emph{anafast} code provided by Healpix software \citep{Gorski05}.
The power spectrum in modes from $l=1$ (dipole) up to $l=20$ was studied. 
Due to statistical and systematic uncertainties higher order modes are irrelevant for this search.
%%%%%%%%%%%%%%%%%%%%%%%%%%%%%%%%%%%%%%%%%%%%%%%%%%%%%%%%%
\begin{figure}[h!]
\centering
\includegraphics[width=10cm]{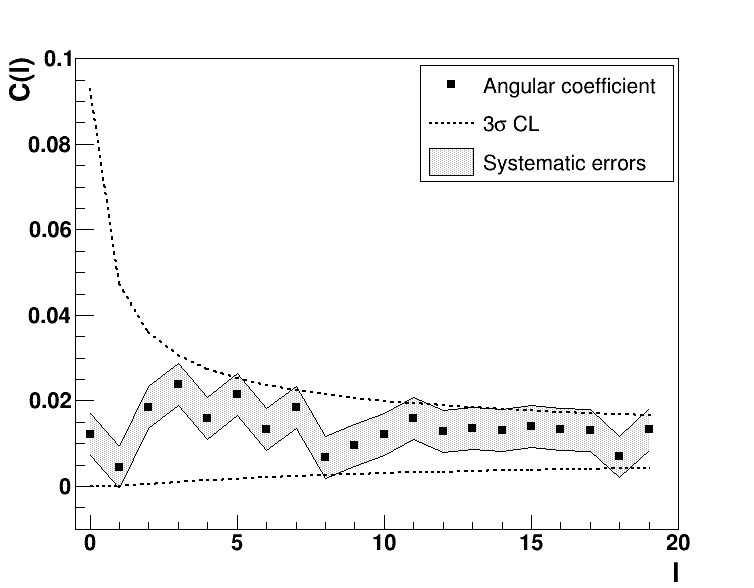}
\caption[Angular power spectra]{Angular power spectra C(l) as a function of the multi-pole l for the positron signal over 
the proton background. The grey area represents the systematic effects calculated as described in Sec.~\ref{sys}.}
\label{fig:AngSpec}
\end{figure}
%%%%%%%%%%%%%%%%%%%%%%%%%%%%%%%%%%%%%%%%%%%%%%%%%%%%%%%%%
In Fig.~\ref{fig:AngSpec} the angular power spectra C(l) as a function of the multi-pole l are shown.
The band defined in the figures by the two dotted lines shows the 3$\sigma$ region of the expected power spectrum from an isotropic sky.
The power spectra are compatible with the prediction of the isotropic sky and no significant deviation from the isotropy was observed.
The value for the measured dipole amplitude $\hat{\delta}_1$, as expressed in Sec. \ref{dip_ul}, is calculated with the value of
coefficient $C_1$ reported in Fig.~\ref{fig:AngSpec}. Solving the integral in Eq. (\ref{eq:ul}) for a confidence level at 95\%,
the limit on the dipole anisotropy parameter is $\delta$ = 0.166.

%%%%%%%%%%%%%%%%%%%%%%%%%%%%%%%%%%%%%%%%%%%%%%%%%%%%%%%%%%%%%%%%%%%%%%%%%%%%%%%%%%%%%%%%%%%%%%%%%%%%%%%%%%%%%%%%%%%%%%%%%%%%%%%%%%%%%%%%%%%%%%%%%%%%%%

\section{Systematic uncertainties estimation}
\label{sys}

Systematic uncertainties were estimated with the following methodology. Variables that have an effect on the anisotropy measurement were 
independently and artificially varied in the selected data samples to understand the effects of the introduced biases on the anisotropy measurement.
For each of the new biased samples of positrons and protons the analysis was performed as explained in the previous sections; the resulting Li and Ma
distribution and angular power spectrum were compared to the un-biased one and the difference was taken as an estimation of systematic
uncertainties for the given bias. Assuming no correlation between the systematic effects (biases), the total
resulting systematic uncertainty was obtained by quadratically summing the single components.

Three main sources of systematic errors were identified:
\begin{enumerate}
\item particle direction reconstruction. Experimentally, the particle direction reconstruction in celestial coordinates results from the combination of the
particle direction in the PAMELA reference frame measured by the tracking system and the pointing direction of the satellite 
(hence of PAMELA which is solidly attached to the Resurs-DK1).
The inclination information of the satellite is provided to PAMELA by the satellite every 1.5-180 s.
The higher rate is used during the satellite movements that are performed to observe the Earth with the optical camera mounted on the satellite. Hence, satellite inclination
information derives from an interpolation of points as function of time and it has a resolution usually better than about two degrees. To estimate the effects of a
possible systematic in the particle direction reconstruction on the anisotropy measurements, the inclination measurement was changed for each event by choosing random values according
to a Gaussian distribution, with a variance equal to the experimental angular resolution.
\item Particle rigidity determination. Particle rigidity is measured by the tracker, by determining the deflection of charged particles in the magnetic field. The maximum
detectable rigidity (MDR) in PAMELA is of about 1.2 TV, resembling the accuracy in the cluster position determination of about 3 $\mu$m in the bending view and the 
spectrometer magnetic field of about 0.4 Tesla. Residual coherent misalignment of the tracker silicon sensors after detector calibration could bring a systematic shift in the rigidity
measurement of about 15\% at the MDR corresponding to a maximum shift in the deflection measurement of about 0.0001~GV$^{-1}$. A systematic error in the
rigidity measurement could have an impact in the anisotropy measurement, since the applied back tracing procedure relies on the particle rigidity to perform the integration of
motion in the Earth's magnetic field. To estimate the effects of these systematic uncertainties, a deflection uncertainty $\pm$10$^{-4}$~GV$^{-1}$ was
added and subtracted to the deflection measurement of selected particles.
\item Event time determination. Absolute UTC time is saved by the PAMELA CPU with a precision of one millisecond. The absolute time of the PAMELA CPU is
synchronized with the satellite with a precision of one second at least once per orbit. Different clock frequencies between the two clocks and the absolute time
at ground are negligible. Time determination is used to calculate the satellite position in latitude, longitude and altitude given the orbital elements. Hence, the
uncertainty in time corresponds to an uncertainty in the orbital positioning and influences the back tracing procedure and the anisotropy estimation. Systematic
effects were estimated by biasing the sample adding and subtracting one second to the measured time on an event-by-event basis.
\end{enumerate}

The resulting estimated systematic uncertainty on the anisotropy measurement is of the order of about 12\%
and shown as grey bands in figures \ref{fig:histoLiMa} and \ref{fig:AngSpec}.
%%%%%%%%%%%%%%%%%%%%%%%%%%%%%%%%%%%%%%%%%%%%%%%%%%%%%%%%%%%%%%%%%%%%%%%%%%%%%%%%%%%%%%%%%%%%%%%%%%%%%%%%%%%%%%%%%%%%%%%%%%%%%%%%%%%%%%%%%%%%%%%%%%%%%%

\section{Search in the direction of the Sun}

An excess of electrons and positrons was also searched for in the Sun
direction. The reference frame used in this analysis is the ecliptic coordinate system centered on the Sun and
referred to the J2000 epoch \citep{Murray89}.

The number of electron and positron events within annuli centered on the Sun in the range 0$^{\circ}$-90$^{\circ}$, in 5$^{\circ}$ steps, is calculated.
%%%%%%%%%%%%%%%%%%%%%%%%%%%%%%%%%%%%%%%%%%%%%%%%%%%%%%%%%
\begin{figure}[h!]
\centering
\includegraphics[width=8cm]{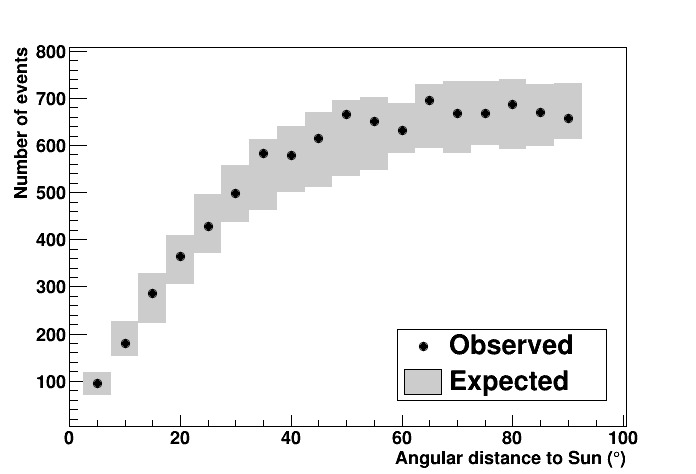}
\caption[Sun]{Total number of electrons and positrons with 10 GV $\leq$ R $\leq$ 200 GV as a function of the angular distance from the Sun.
The grey boxes correspond to the 3$\sigma$ fluctuation respect to an isotropic expectation.}
\label{fig:sun}
\end{figure}
%%%%%%%%%%%%%%%%%%%%%%%%%%%%%%%%%%%%%%%%%%%%%%%%%%%%%%%%%
The results are reported in Fig.~\ref{fig:sun} as a function of the angular distance from the Sun direction
and are compared to the expectations from an isotropic sky. The grey boxes represent the 3$\sigma$ bounds for an
isotropic sky calculated with an elliptical reference frame on subsets of coverage map. Data are consistent with the isotropic expectation within a 3$\sigma$ interval. 
Similar results were obtained by Fermi LAT collaboration \citep{Ajello11} without selecting electrons and positrons separately.
%%%%%%%%%%%%%%%%%%%%%%%%%%%%%%%%%%%%%%%%%%%%%%%%%%%%%%%%%%%%%%%%%%%%%%%%%%%%%%%%%%%%%%%%%%%%%%%%%%%%%%%%%%%%%%%%%%%%%%%%%%%%%%%%%%%%%%%%%%%%%%%%%%%%%%

\section{Discussion}
\label{disc}

The detection of an anisotropy signal might be used to distinguish
between models that could explain the cosmic-ray positron excess, 
e.g. \citep{Adriani09}, 
over the predictions of commonly used propagation models. 
A variety of models, invoking either dark
matter or astrophysical objects as 
positron sources, e.g. see review papers \citep{Boezio09,Serpico12}, 
have been put 
forward to explain this excess. As a result
of the different localization of such sources an anisotropy in the
arrival particle distribution cannot be excluded.
Various authors,
e.g. \citep{Ptuskin95,Ackermann10,DiBernardo11,Kobayashi04,Bushing08},
studied the  
expected anisotropy in cosmic-ray positrons from supernova 
remnants and/or from pulsars. For example, 
\citet{Ackermann10}, \citet{DiBernardo11} estimated
an anisotropy of the order of about 1\% for 
100 GeV positrons accelerated by Vela supernova 
remnant,
while \citet{Bushing08} estimated an anisotropy of 
about 3\% and about 0.3\% for positrons with energies between 10 and
100 GeV 
accelerated by Monogem
and Geminga pulsars, respectively. 
In the case a dark matter origin, \citet{Ackermann10} estimated 
an anisotropy 
of the order of per-mil. 

A blind search for positron anisotropies was performed on PAMELA data,
taking into account the effects of the Earth's geomagnetic field. 
The analysis of PAMELA data was carried out by using two techniques:
the significance 
maps, which allowed the study of the directions 
of possible over-densities at any angular scale, not just 
the dipole contribution, and the angular power  
spectrum, which is a method capable of determining the anisotropy magnitude.
Results were consistent with isotropy at all angular scales considered.
A limit on the dipole anisotropy parameter, $\delta$ = 0.166 at 95\%
confidence level, was obtained. This limit is close but higher than
the expected anisotropy for astrophysical sources and about an order
of magnitude higher than in dark matter models that, therefore, cannot be
excluded. 
Because of the limited statistics of the positron data set it was not
possible to investigate the energy
dependence of the dipole amplitude. 

A few models \citep{Batell10,Schuster10} predict 
that large astrophysical bodies within the Solar
System may significantly contribute to 
cosmic-ray positrons capturing dark matter particles that 
may annihilate first into metastable mediators.
If the mean free path of these mediators is in excess of the solar (or
planetary) radius, they can escape and annihilate again, 
producing gamma rays and charged particles such as positrons
arriving from 
directions correlated with the centers of the astrophysical 
source. The analysis of PAMELA all electron data in the direction of the Sun are
consistent with the isotropic expectation at a 99.7\% confidence level.

Results of the analysis reported in this paper are in agreement with those published by AMS
and Fermi LAT. It should be noted, however, that a special care 
is needed when comparing the results of these three different
experiments. AMS results \citep{Accardo14} refer to positrons but 
reporting only the dipole upper limit, 
while in this paper a multipole analysis is presented. 
Fermi LAT results \citep{Ackermann12} refer to the sum of electrons
and positrons. 
In the energy range between tens of GeV and hundreds of GeV,
electrons from standard cosmic-rays sources are the 
dominant component. 
Hence, the study of the pure positron signal is more sensitive, in
this energy range, to anisotropies. 
Finally, in this analysis the effects of the Earth's magnetic field,
which could affect a weak anisotropy,
were taken into account, while both the AMS 
and Fermi LAT results refer to the anisotropy as measured in a low
Earth orbit. 

It should be noted that the local (a few tens of parsec) turbulent
realization of the interstellar magnetic
field \citep{Giacinti12} as well as magnetic reconnection in the
heliotail \citep{Lazarian10} may have an effect on 
the cosmic-ray arrival directions. Such processes have been proposed
to explain the anisotopy observed mostly in the multi-TeV energy
region
\citep{Nagashima98,Amenomori06,Guillian07,Abdo09,Aglietta09,Abbasi10,Iuppa13}.
However, there are no quantitative studies on the effects of these
processes on the arrival directions of tens of GeV positrons measured
at Earth.

%%%%%%%%%%%%%%%%%%%%%%%%%%%%%%%%%%%%%%%%%%%%%%%%%%%%%%%%%%%%%%%%%%%%%%%%%%%%%%%%%%%%%%%%%%%%%%%%%%%%%%%%%%%%%%%%%%%%%%%%%%%%%%%%%%%%%%%%%%%%%%%%%%%%%%
\section{Conclusions}

The arrival direction of cosmic-ray positrons in the rigidity range
from 10 GV to 200 GV for 
about 10$^3$ positrons collected in the first 4 years of operation of
PAMELA was studied. The effects of the Earth's geomagnetic field were
accounted for. Results are 
consistent with isotropy at all angular scales considered. Moreover, no
enhancement of the all electron  
flux was observed in the direction towards the Sun.

%%%%%%%%%%%%%%%%%%%%%%%%%%%%%%%%%%%%%%%%%%%%%%%%%%%%%%%%%%%%%%%%%%%%%%%%%%%%%%%%%%%%%%%%%%%%%%%%%%%%%%%%%%%%%%%%%%%%%%%%%%%%%%%%%%%%%%
%%%%%%%%%%%%%%%%%%%%%%%%%%%%%%%%%%%%%%%%%%%%%%%%%%%%%%%%%%%%%%%%%%%%%%%%%%%%%%%%%%%%%%%%%%%%%%%%%%%%%%%%%%%%%%%%%%%%%%%%%%%%%%%%%%%%%%%%%%%%%%%%%%%%%%
\section*{Acknowledgments}
	
We would like to acknowledge contributions and support from: our Universities and Institutes, The Italian Space Agency (ASI),
Deutsches Zentrum f\"ur Luft - und Raumfahrt (DLR), The Swedish National Space Board, The Swedish Research Council,
The Russian Space Agency (Roscosmos) and Russian Scientific Foundation.

%%%%%%%%%%%%%%%%%%%%%%%%%%%%%%%%%%%%%%%%%%%%%%%%%%%%%%%%%%%%%%%%%%%%%%%%%%%%%%%%%%%%%%%%%%%%%%%%%%%%%%%%%%%%%%%%%%%%%%%%%%%%%%%%%%%%%%%%%%%%%%%%%%%%%%


\begin{thebibliography}{}

\bibitem[Abbasi et al.(2010)]{Abbasi10}
Abbasi, R. et~al. 2010, ApJ., 718, L194
\bibitem[Abdo et al.(2009)]{Abdo09}
Abdo, A. A. et al. 2009,ApJ, 698, 2121
\bibitem[Accardo et al.(2014)]{Accardo14}
 Accardo, L., Aguilar, M. \& Aisa, D. et al. 2014, PhRvL, 113, 121101
\bibitem[Ackermann et al.(2010)]{Ackermann10}
 Ackermann, M., Ajello, M., \& Atwood, W. B. et al. 2010, PhRvD, 82, 092003
\bibitem[Ackermann et al.(2012)]{Ackermann12}
 Ackermann, M., Ajello, M., \& Allafort, A. et al. 2012, PhRvL, 108, 011103
\bibitem[Adriani et al.(2009)]{Adriani09}
 Adriani, O., Barbarino, G. C., \& Bazilevskaya, G. A. et al. 2009, Nature, 458, 486
\bibitem[Adriani et al.(2010)]{Adriani10}
 Adriani, O., Barbarino, G. C., \& Bazilevskaya, G. A. et al. 2010, APh, 34, 111
\bibitem[Adriani et al.(2011)]{Adriani11}
 Adriani, O., Barbarino, G. C., \& Bazilevskaya, G. A. et al. 2011,
PhRvL, 106, 201101
\bibitem[Adriani et al.(2013)]{Adriani13}
 Adriani, O. et al. 2013, PhRvL, 111, 081102
\bibitem[Aglietta et al.(2009)]{Aglietta09}
Aglietta, M. et al. 2009,ApJ, 692, L130
\bibitem[Ajello et al.(2011)]{Ajello11}
 Ajello, M., Atwood, W. B., \& Baldini, L. et al. 2011, PhRvD, 84, 032007
\bibitem[Amenomori et al.(2006)]{Amenomori06}
Amenomori, M. et al. 2006, Science, 314, 439
\bibitem[Batell et al.(2010)]{Batell10}
 Batell, B., Pospelov M., Ritz, A., \& Shang, Y. 2010, PhRvD, 81, 075004
\bibitem[Blasi(2009)]{Blasi09}
 Blasi, P. 2009, PRL 103, 051104
\bibitem[Boezio et al.(2009)]{Boezio09}
Boezio, M. et~al. 2009, New\ J.\ Phys., 11, 105023.
\bibitem[Bruno et al.(2014)]{Bruno14}
 Bruno, A., Adriani, O., \& Barbarino, G. C. et al. 2014, arXiv:1412.1765
\bibitem[B\"ushing et al.(2008)]{Bushing08}
 B\"usching, I., de Jager, O.C., Potgieter, M.S., \& Venter, C. 2008, ApJ, 678, L39
\bibitem[Campana et al.(2013)]{Campana13}
 Campana, D., Giaccari, U., \&, Adriani, O. et al. 2013, JPhCS, 409,
012055
\bibitem[Cirelli et al.(2008)]{Cirelli08}
Cirelli, M., Kadastik, M., Raidal, M., \& Strumia, A. 2008, Nucl.\
Phys.\ B, 813, 1
% \bibitem[Casaus(2013)]{Casaus13}
%  Casaus, J., AMS collaboration 2013, Proceedings 33rd ICRC, Rio de Janeiro
\bibitem[Delahaye et al.(2009)]{Delahaye09}
 Delahaye, T., Lineros, R., \& Donato, F. et al. 2009, A\&A, 501, 821
\bibitem[Desiati \& Lazarian(2013)]{Desiati13}
 Desiati, P., \& Lazarian, A. 2013, ApJ, 762, 44
\bibitem[Di Bernardo et al.(2011)]{DiBernardo11}
 Di Bernardo, G., Evoli, C., Gaggero, D. et al. 2011, APh, 34, 528
\bibitem[Finlay et al.(2010)]{Finlay10}
 Finlay, C. C., Maus, S., \& Begganet, C. D. et al. 2010, GeoJI, 183, 1216
\bibitem[Garthwaite et al.(1995)]{Garthwaite95} 
Garthwaite, P. H., Jolliffe, I. T. \& Jones, B. 1995, Statistical Inference (Prentice Hall)
\bibitem[Giaccari et al.(2013)]{Giaccari13}
 Giaccari, U., Adriani, O., \& Barbarino, G. C., et al. 2013, NuPhS, 239, 123
\bibitem[Gorski et al.(2005)]{Gorski05}
 Gorski, K. M., Hivon, E., \& Banday, A. J. et al. 2005, ApJ, 622, 759
\bibitem[Grasso et al.(2009)]{Grasso09}
 Grasso, D., Profumo, S., \& Strong, A.W. et al. 2009, APh, 32, 140
\bibitem[Giacinti et al.(2012)]{Giacinti12}
 Giacinti, G., \& Sigl, G. 2012, Phys. Rev. Lett., 109, 071101
\bibitem[Guillian et al.(2007)]{Guillian07}
Giullian, G. et al. 2007, Phys. Rev. D, 78, 062003
\bibitem[Hooper et al.(2009)]{Hooper09}
 Hooper, D., Blasi, P., \& Serpico, D. 2009, JCAP, 0901, 025
\bibitem[Iuppa et al.(2013)]{Iuppa13}
Iuppa, R. et al. 2013, J. Phys.: Conf. Ser., 409, 012039
\bibitem[Kobayashi et al.(2004)]{Kobayashi04}
 Kobayashi, T., Komori, Y., Yoshida K., \& Nishimura, J. 2004, ApJ, 601, 340
\bibitem[Lazarian et al.(2010)]{Lazarian10}
Lazarian, A., \& Desiati, P. (2010), ApJ, 722, 188
\bibitem[Li et al.(1983)]{Li83}
 Li, T.P., \& Ma, Y.Q. 1983, ApJ, 272, 317
\bibitem[Moskalenko \& Strong (2009)]{Moskalenko09}
 Moskalenko, I. V., \& Strong, A. W. 1998, ApJ, 493, 694
\bibitem[Murray(1989)]{Murray89}
 Murray, C. A. 1989, A\&A, 218, 325
\bibitem[Nagashima et al.(1998)]{Nagashima98}
Nagashima, K. et al. 1998, J. of Geoph. Res., 103, 17429
\bibitem[Neyman(1937)]{Neyman37} 
 Neyman, J., 1937, Philosophical Transactions of the Royal Society of London. Series A, Math. and Phys. Sciences, 236, 767
\bibitem[Picozza et al.(2007)]{Picozza07}
 Picozza, P. et al. 2007, APh, 27, 296
\bibitem[Ptuskin et al.(1995)]{Ptuskin95}
 Ptuskin, V. S., \& Ormes, J. F. 1995, Proceedings 24th ICRC, Rome
\bibitem[Schuster et al.(2010)]{Schuster10}
 Schuster, P., Toro, N., \& Yavin, I. 2010, PhRvD, 81, 016002
\bibitem[Serpico(2012)]{Serpico12}
Serpico, P. 2012, Astropart.\ Phys., 39, 2
\bibitem[Smart et al.(2000)]{Smart00}
 Smart, D. F., \& Shea, M. A. 2000, Final Report, Grant NAG5-8009, Center for Space Plasmas and Aeronomic Research, The University of Alabama in Huntsville.
\bibitem[Tsyganenko et al.(2005)]{Tsyganenko05}
Tsyganenko, N. A., \& Sitnov, M. I. 2005, JGR, 110, A03208
\end{thebibliography}
\end{document}